\documentclass[pra,showpacs,floatfix,twocolumn,superscriptaddress]{revtex4}
\usepackage{graphicx,amsmath,amssymb}
\usepackage{epsf,color}
\usepackage{pslatex,floatflt,latexsym}

\begin{document}
\title{Strong Violations of Bell-type Inequalities for Path-Entangled Number States}
\author{Christoph~F.~Wildfeuer}
\email{wildfeuer@phys.lsu.edu}
\affiliation{Hearne Institute for Theoretical Physics, Department of Physics
  and Astronomy, Louisiana State University,
 Baton Rouge, Louisiana 70803, USA}
\author{Austin P. Lund}
\affiliation{Centre for Quantum Computer Technology, Department of Physics, University of Queensland, 
 Brisbane 4072, QLD, Australia}
\author{Jonathan~P.~Dowling} 
\affiliation{Hearne Institute for Theoretical Physics, Department of Physics
  and Astronomy, Louisiana State University, 
 Baton Rouge, Louisiana 70803, USA}
\begin{abstract}
We show that nonlocal correlation experiments on the two spatially separated
modes of a maximally path-entangled number state may be performed. They lead to a violation of a
Clauser-Horne Bell inequality for any finite photon number $N$. We also present an analytical expression for the two-mode Wigner function of a maximally path-entangled number state and investigate a Clauser-Horne-Shimony-Holt Bell
inequality for such a state. We test other Bell-type inequalities. Some are violated by a constant amount for any $N$.
\end{abstract}
\maketitle
\section{Introduction}\label{Intro}
\par Maximally path-entangled number states of the form  
\begin{eqnarray}\label{NOON}
  |\Psi\rangle=\frac{1}{\sqrt{2}}(|N\rangle_a|0\rangle_b+e^{i\varphi}|0\rangle_a|N\rangle_b)\,,
\end{eqnarray}
(often referred as N00N states) have im\-por\-tant applications to quan\-tum imaging \cite{Boto}, metrology
\cite{Lee,Migdall}, and sensing \cite{Lee2}. Characterizing their quantum
mechanical properties is therefore a valuable task for improving upon the
above applications. Entanglement is the most
profound property of quantum mechanical systems. N00N states are non-separable states
and hence are entangled. But do they also show nonlocal
behavior when we perform a correlation experiment on their modes? The amount
of nonlocality demonstrated by a Bell-type experiment provides an
operational definition of entanglement (for a review of Bell inequalities and
experiments see, e.g., \cite{Genovese}. It distinguishes between the class of
states that are entangled but admit a local hidden variable model and those
which do not and so may be called EPR correlated \cite{Werner}. 
\par Several publications \cite{Reed} address the
question of whether the N00N states are EPR correlated for the case
$N=1$. Gisin and Peres have shown that it is
possible to find pairs of observables, whose correlations violate a Bell's
inequality for any nonfactorable pure state of two quantum systems
\cite{Gisin}. This result was later extended to states of more than two systems by Popescu and Rohrlich \cite{Popescu}. Recent experiments
\cite{Lvovsky,Bellini} have reported strong evidence that N00N states
violate a Bell's inequality for $N=1$, leaving open the question as to what
experiments might show EPR correlations for $N>1$. 
\par We propose a specific experiment that shows that N00N states are EPR correlated for any finite $N$. We investigate two
measurement schemes using the unbalanced homodyne detection scheme described
in \cite{Banaszek} and compare the results. The correlation functions we
calculate can be related to two well-known phase space distributions: the two-mode $Q$ function and
the two-mode Wigner function. Banaszek
and W\'odkiewicz first pointed out the operational definition of
the $Q$ and Wigner function \cite{Banaszek}. We modify this approach
and calculate the distribution functions for the N00N states entirely from
these phase space distributions, and thereby construct a Clauser-Horne and a
Clauser-Horne-Shimony-Holt Bell inequality. In section \ref{new_inequalities} we also test other
Bell-type inequalities not commonly used so far in quantum optical experiments.
\par
\begin{figure}[htb]
\includegraphics[scale=0.38]{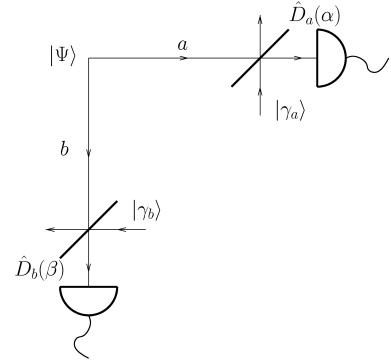}
\caption{Unbalanced homodyne detection scheme for a Bell experiment with N00N
  states. Here $|\Psi\rangle=\frac{1}{\sqrt{2}}(|N\rangle_a|0\rangle_b-|0\rangle_a|N\rangle_b)$
  and $a$ and $b$ label the modes.\label{homodyne}}
\end{figure}

 The above Bell-tests may be performed in an unbalanced homodyne detection
scheme as given, for example, in Ref.\cite{Banaszek} and shown in
Fig.\ \ref{homodyne}. 
For simplicity we choose $\varphi=\pi$ for the states in Eq.~(\ref{NOON}). 
\par It is now understood that the introduction of a reference frame is
required in any Bell test \cite{Bartlet} and one should consider the
field modes as entangled rather than the photons \cite{vanEnk,vanEnk_2}. In the number basis, a shared local oscillator acts as the required reference frame. 
The beam splitters in this approach are assumed to operate in the limit where the transmittivity $T\rightarrow 1$.
We further assume that a strong coherent state $|\gamma\rangle$, where
$|\gamma|\rightarrow\infty$, is incident on one of the two input ports. The beam
splitter then acts as the displacement operator $\hat{D}(\gamma\sqrt{1-T})$ on
the second input port \cite{Wiseman,Banaszek_2,Wallentowitz}. We
introduce complex parameters $\alpha=\gamma_a\sqrt{1-T}$ and $\beta=\gamma_b\sqrt{1-T}$. The phase-space parameterization with respect
to these parameters is then analogous to a correlation experiment with polarized light
and different relative polarizer settings where the nonlocality of
polarization entangled states such as
$|\Psi\rangle=(|H\rangle_a|V\rangle_b-|V\rangle_a|H\rangle_b)/\sqrt{2}$ is
well established. 
\section{Bell experiment with on-off detection scheme}\label{on/off}
\par
\begin{figure}[htb]
     \setlength{\unitlength}{1cm}  
         \begin{minipage}[t]{6cm}
         \begin{picture}(6,4)\includegraphics[scale=0.55]{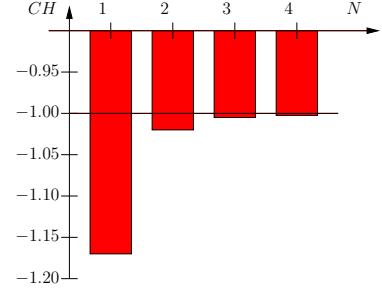}\end{picture}
           \caption{Violation of the Clauser-Horne Bell inequality as a function
             of $N$.}\label{figure1}
          \end{minipage}
          \end{figure}

 In the first experimental setup we consider a simple nonnumber resolving
photon-detection scheme. In the case of the homodyne detection scheme under
consideration, the local positive operator valued measure (POVM) is
defined by $\hat{Q}(\alpha)+\hat{P}(\alpha)=\hat{{\bf 1}}$ with,  
\begin{eqnarray}
\hat{Q}(\alpha)&=&\hat{D}(\alpha)|0\rangle\langle 0|\hat{D}^\dagger(\alpha)\,,\label{povm1}\\
\hat{P}(\alpha)&=&\hat{D}(\alpha)\sum_{n=1}^\infty|n\rangle\langle
n|\hat{D}^\dagger(\alpha)\,.\label{povm2}
\end{eqnarray}
We assume lossless detectors for our investigation.
The expectation value of $\hat{Q}(\alpha)$ tells us the probability that no
photons are present, depending on the phase and amplitude of the local
oscillator. The expectation value of $\hat{P}(\alpha)$ gives the
probability of counting one or more photons, while not distinguishing between
one or more photons. So we simply assign a 1 to a detector click and a 0 otherwise, giving us a binary result.
We label the two modes of the N00N state by $a$ and $b$. The corresponding
measurement operators for a correlated measurement of the displaced vacuum can
be written as $\hat{Q}_a(\alpha)\otimes\hat{Q}_b(\beta)$. The expectation value for the state $|\Psi\rangle$ is given by 
 \begin{eqnarray}\label{Qfunction}
   Q_{ab}(\alpha,\beta)=\langle\Psi|\hat{Q}_a(\alpha)\otimes\hat{Q}_b(\beta)|\Psi\rangle=|\langle
   \alpha,\beta|\Psi\rangle|^2\,.
\end{eqnarray} 
The above expression is the two-mode $Q$ function of the N00N state up to a factor $1/\pi^2$,
and the result is given by 
\begin{eqnarray}
  Q_{ab}(\alpha,\beta)=\frac{1}{2N!}e^{-(|\alpha|^2+|\beta|^2)}|\alpha^N-\beta^N|^2\,.\label{Qfunction1}
\end{eqnarray}
To obtain the probabilities for the individual measurements we calculate 
\begin{eqnarray}
  Q_{a}(\alpha)&=&\langle\Psi|\hat{Q}_{a}(\alpha)\otimes \hat{{\bf
  1}}_{b}|\Psi\rangle=\frac{1}{2}{
  e}^{-|\alpha|^2}\left(\frac{|\alpha|^{2N}}{N!} +1\right),\label{Qfunction2}\\
 Q_{b}(\beta)&=&\langle\Psi|\hat{{\bf
  1}}_{a}\otimes\hat{Q}_{b}(\beta)|\Psi\rangle=\frac{1}{2}{
  e}^{-|\beta|^2}\left(\frac{|\beta|^{2N}}{N!} +1\right).\label{Qfunction3}
\end{eqnarray}
Using the completeness relation $\hat{P}(\alpha)=\hat{{\bf
    1}}-\hat{Q}(\alpha)$, we obtain the probabilities for the
    correlated and single detector counts
    --- $P_{a}(\alpha)=1-Q_a(\alpha)$, $P_{b}(\beta)=1-Q_b(\beta)$, and $P_{ab}(\alpha,\beta)=1-Q_a(\alpha)-Q_b(\beta)+Q_{ab}(\alpha,\beta)$ --- in terms of the $Q$ functions.  
We build from these the Clauser-Horne
combination (CH) \cite{Clauser-Horne}, which for a local hidden variable model admits the
inequality,
\begin{eqnarray}
  -1&\le& P_{ab}(\alpha,\beta)-P_{ab}(\alpha,\beta')+P_{ab}(\alpha',\beta)+P_{ab}(\alpha',\beta')\nonumber\\
   & & -P_{a}(\alpha')-P_{b}(\beta)\le 0\,.\label{CH}
\end{eqnarray}
If this inequality is violated it follows that N00N states contain nonlocal
correlations, i.e., EPR correlations. In order to attain such a violation, we
minimize the function
$CH=P_{ab}(\alpha,\beta)-P_{ab}(\alpha,\beta')+P_{ab}(\alpha',\beta)+P_{ab}(\alpha',\beta')-P_{a}(\alpha')-P_{b}(\beta)$
for a given $N$ over the parameter space spanned by $\alpha$, $\alpha'$, $\beta$, and $\beta'$.
The violation of the Clauser-Horne combination for the N00N states with $N=1,\ldots,4$ is
shown in Fig.\ \ref{figure1}.  
The results show a decrease in the amount of violation with $N$. The maximal
violation is obtained for $N=1$. For $N\ge 3$ the violation is so reduced that
it would be increasingly hard to observe experimentally. If we increase the
precision of our numerical method, we observe that for large $N$ the minimum of
the CH combination, in fact, never hits the classical bound of $-1$ exactly, i.e., there is a
violation of the inequality for any finite $N$, which can be shown
as follows. Let $N$ be fi\-nite and odd. We choose $\alpha'=\beta=0$ and $\alpha=-\beta'$, then the CH combination
reduces to $CH=1/N!\,|\alpha|^{2N}e^{-|\alpha|^2}(1-2e^{-|\alpha|^2})-1$. For
  any $0<|\alpha|^2<\ln 2$, we obtain $CH<-1$. For even $N$ the same proof holds
  except that we need to choose $\alpha=\beta'$ instead.
\par The Bell measurement presented leads to a decrease of the amount of
         violation with $N$. This decrease with $N$ is due to the specific way the reference
         frame is introduced in terms of the local displacement
         operators $\hat{D}(\alpha)$ and $\hat{D}(\beta)$ for the correlation measurement.
         The scheme is based on measuring the overlap of coherent states
         with the modes of the N00N state. The elements contained in
         Eq.~(\ref{Qfunction}) are of the form $\langle
         N,0|\alpha,\beta\rangle$ and $\langle 0,N|\alpha,\beta\rangle$. In
         order to maximize those products we would need $\alpha$ to take, at the same time, the values
         $|\alpha|^2=N$ and $|\alpha|^2=0$. Since the `distance' of $N$ to the vacuum becomes
         larger with $N$, the correlated overlap is reduced. This may explain the decrease in
         the amount of violation observed. 
         \par
\begin{figure}[htb]
     \setlength{\unitlength}{1mm}
     \begin{picture}(45,45)
     \put(0,0){\includegraphics[scale=0.4]{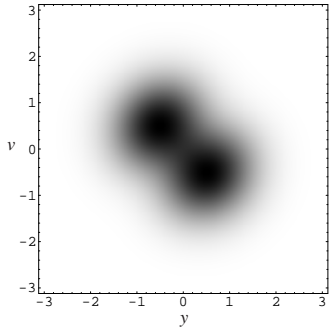}}
     \put(-2,21){$\scriptstyle v$}
     \put(21,-2){$\scriptstyle y$}
   \end{picture}      
     \caption{The marginal $Q$ function $Q_\mathrm{m}(y,v)$ for $N=1$.}\label{plotQ_1}
       \end{figure}

\begin{figure}[htb]
     \setlength{\unitlength}{1mm}
     \begin{picture}(45,45)
     \put(0,0){\includegraphics[scale=0.4]{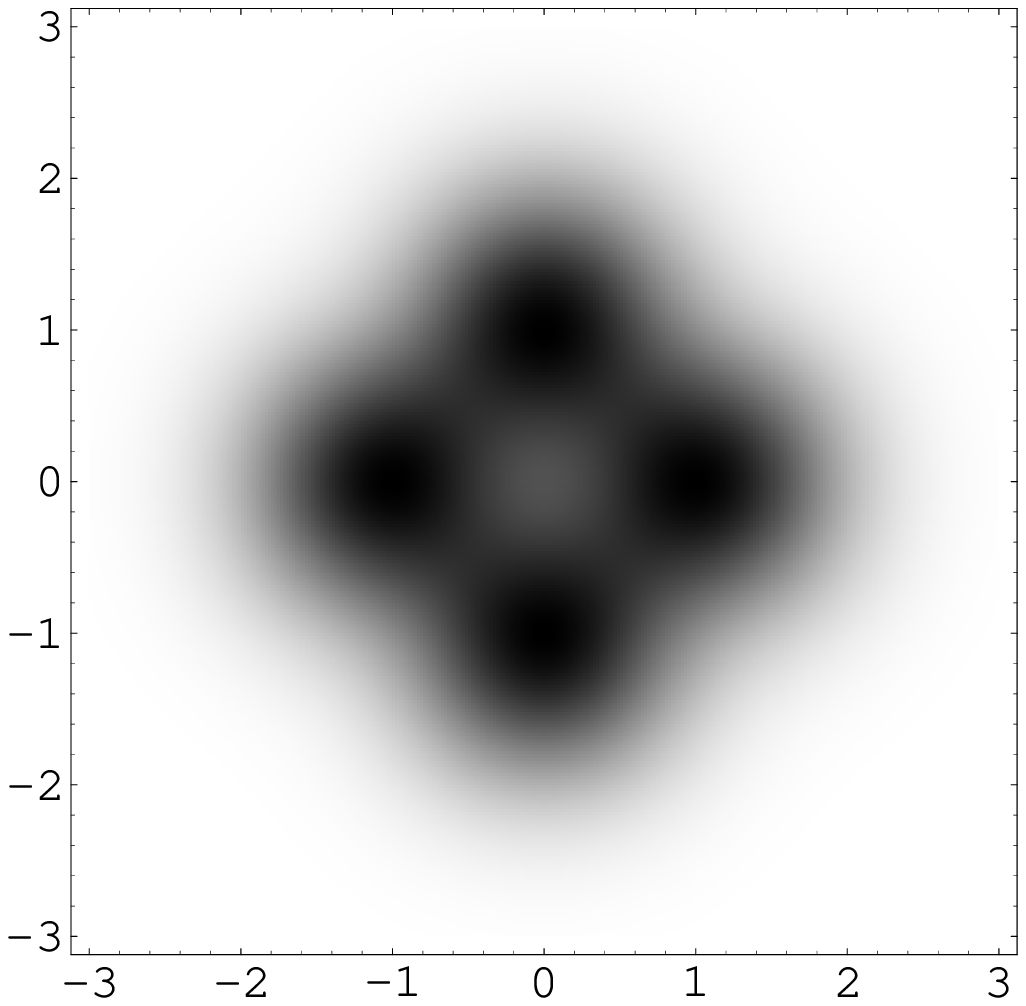}}
     \put(-2,21){$\scriptstyle v$}
     \put(21,-2){$\scriptstyle y$}
   \end{picture}      
     \caption{The marginal $Q$ function $Q_\mathrm{m}(y,v)$ for $N=2$.}\label{plotQ_2}
       \end{figure}

\begin{figure}[htb]
     \setlength{\unitlength}{1mm}
     \begin{picture}(45,45)
     \put(0,0){\includegraphics[scale=0.4]{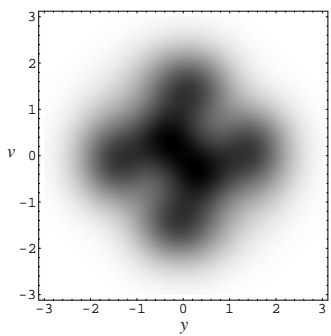}}
     \put(-2,21){$\scriptstyle v$}
     \put(21,-2){$\scriptstyle y$}
   \end{picture}      
     \caption{The marginal $Q$ function $Q_\mathrm{m}(y,v)$ for $N=3$.}\label{plotQ_3}
       \end{figure}

 We can also display some correlations by plotting the marginals of the $Q$ function in
         Eq.~(\ref{Qfunction}). We therefore decompose the dimensionless
         complex local oscillator amplitudes in the set of real variables $x,y,u,v$, i.e., $\alpha=x+i\, y$,
         $\beta=u+i\, v$, and obtain $Q_\mathrm{m}(y,v)=\int_{-\infty}^\infty\int_{-\infty}^{\infty}Q_{ab}(x,y,u,v)\,dx\,du$.
         These probability densities are displayed in Fig.~\ref{plotQ_1},
         \ref{plotQ_2}, and \ref{plotQ_3}, for $N=1,2,3$. We see that the
         distributions for $N=2,3$ have a higher symmetry than for $N=1$. 
     \par 
\par 
 \par         
         The linear correlation coefficient $r=\mathrm{cov}(y,v)/(\Delta y\Delta v)$,
         where $\mathrm{cov}(y,v)=\int_{-\infty}^\infty\int_{-\infty}^\infty
         (y-\bar{y})(v-\bar{v})Q_\mathrm{m}(y,v)\,\,dy\,\, dv$, 
         vanishes for all $N>1$, although we see from the pictures that the two phase space variables are statistically dependent. This is an indication of nonlinear
         correlations between the two phase space variables. 
Note that the measurement described by
the operators in Eq.~(\ref{povm1}) and Eq.~(\ref{povm2}) requires only non-number resolving
photon counters and may therefore be performed with current detector
technology. In the next section we consider a correlated parity measurement on the modes and investigate the amount of violation in this scheme.
\section{Bell test with parity measurement}\label{parity}
\par An operational definition of the two-mode Wigner function for the N00N state is
given in terms of a correlated parity measurement \cite{Banaszek}. The measurements can be described by the
following POVM operators:
\begin{eqnarray}
 \hat{\Pi}^{+}(\alpha)&=&\hat{D}(\alpha)\sum_{k=0}^\infty|2k\rangle\langle2k|\hat{D}^\dagger(\alpha)\,,\\
 \hat{\Pi}^{-}(\alpha)&=&\hat{D}(\alpha)\sum_{k=0}^\infty|2k+1\rangle\langle 2k+1|\hat{D}^\dagger(\alpha)\,.
\end{eqnarray}
The corresponding operator for the correlated measurement of the parity on
mode $a$ and $b$ may be defined as:
\begin{eqnarray*}
  \hat{\Pi}(\alpha,\beta)=\left(\hat{\Pi}_a^{(+)}(\alpha)-\hat{\Pi}_a^{(-)}(\alpha)\right)\otimes\left(\hat{\Pi}_b^{(+)}(\beta)-\hat{\Pi}_b^{(-)}(\beta)\right).
\end{eqnarray*}
The outcome of the measurements are either $+1$ or $-1$.
It may be noted that this operator can be rewritten as
\begin{eqnarray}\label{Wigner}
  \hat{\Pi}(\alpha,\beta)=\hat{D}_a(\alpha)\hat{D}_b(\beta)(-1)^{\hat{n}_a+\hat{n}_b}\hat{D}_a^\dagger(\alpha)\hat{D}_b^\dagger(\beta)\,,
\end{eqnarray}
and is equivalent to the operator for the Wigner function in
\cite{Royer,Moya} (up to a factor $4/\pi^2$). We note that the operator in
Eq.~(\ref{Wigner}) is essentially a product of operators for mode $a$ and $b$: 
\begin{eqnarray}\label{Wignerproduct}
  \hat{\Pi}(\alpha,\beta)=\hat{D}_a(\alpha)(-1)^{\hat{n}_a}\hat{D}_a^\dagger(\alpha)\hat{D}_b(\beta)(-1)^{\hat{n}_b}\hat{D}_b^\dagger(\beta)\,.
\end{eqnarray}
Using this property the expectation value of Eq.~(\ref{Wignerproduct}) for the
N00N state can be expressed as a function of two Laguerre polynomials
and an interference term,
\begin{eqnarray}\label{wigner}
  \lefteqn{\Pi(\alpha,\beta)=\langle\Psi|\hat{\Pi}(\alpha,\beta)|\Psi\rangle=}\nonumber\\ 
 &&\frac{1}{2}e^{-2|\alpha|^2-2|\beta|^2}\large[(-1)^N(L_N(4|\alpha|^2)+L_N(4|\beta|^2))\nonumber\\
 &&-\frac{2^{2N}}{N!}({\alpha^*}^N\beta^N+\alpha^N{\beta^*}^N)\large]\,,
\end{eqnarray}
where $L_N(x)$ is the Laguerre polynomial \cite{Stegun}. The two-mode
 Wigner function is obtained from
 $W(\alpha,\beta)=\Pi(\alpha,\beta)4/\pi^2$. By building the CHSH
 \cite{clauser} inequality defined by
 \begin{eqnarray}
   -2\le\Pi(\alpha,\beta)+\Pi(\alpha',\beta)+\Pi(\alpha,\beta')-\Pi(\alpha',\beta')\le 2\,,\label{CHSH}
  \end{eqnarray} 
we determine how this Bell inequality is violated as a function of $N$.
A minimization procedure in the parameter space $\alpha$, $\alpha'$, $\beta$, and $\beta'$ as a
 function of $N$ is carried out with a numerical routine to investigate the amount of
 violation. We see that the correlated parity measurement leads to a
         vio\-lation of the CHSH Bell inequality for $N=1$, and that states with larger $N$ do
         not violate the inequality.
 \par The Wigner function may also be
         used to understand this
         behavior.  We therefore calculate the
         marginals of the Wigner function by integrating over two of the
         variables, where we use the same decomposition of the dimensionless
         complex local oscillator amplitudes $\alpha=x+i\, y$ and
         $\alpha=u+i\, v$, and obtain 
         $W_\mathrm{m}(y,v)=\int_{-\infty}^{\infty}\int_{-\infty}^{\infty}W(x,y,u,v)dx\,du$.
         The function $W_\mathrm{m}(y,v)$ is positive definite and can be interpreted
         as the probability density for the remaining variables. 
\begin{figure}[htb]
     \setlength{\unitlength}{1mm}
     \begin{picture}(40,40)
     \put(0,0){\includegraphics[scale=0.4]{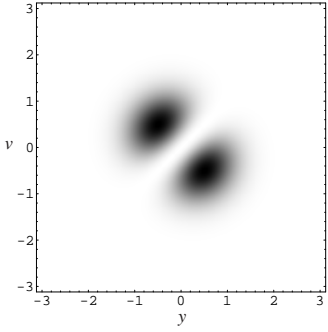}}
     \put(-2,21){$\scriptstyle v$}
     \put(21,-2){$\scriptstyle y$}
   \end{picture}      
     \caption{The marginal Wigner function $W_\mathrm{m}(y,v)$ for $N=1$.}\label{Wigner1}
       \end{figure} 
    \begin{figure}[htb]
      \setlength{\unitlength}{1mm}
     \begin{picture}(40,40) 
      \put(0,0){\includegraphics[scale=0.4]{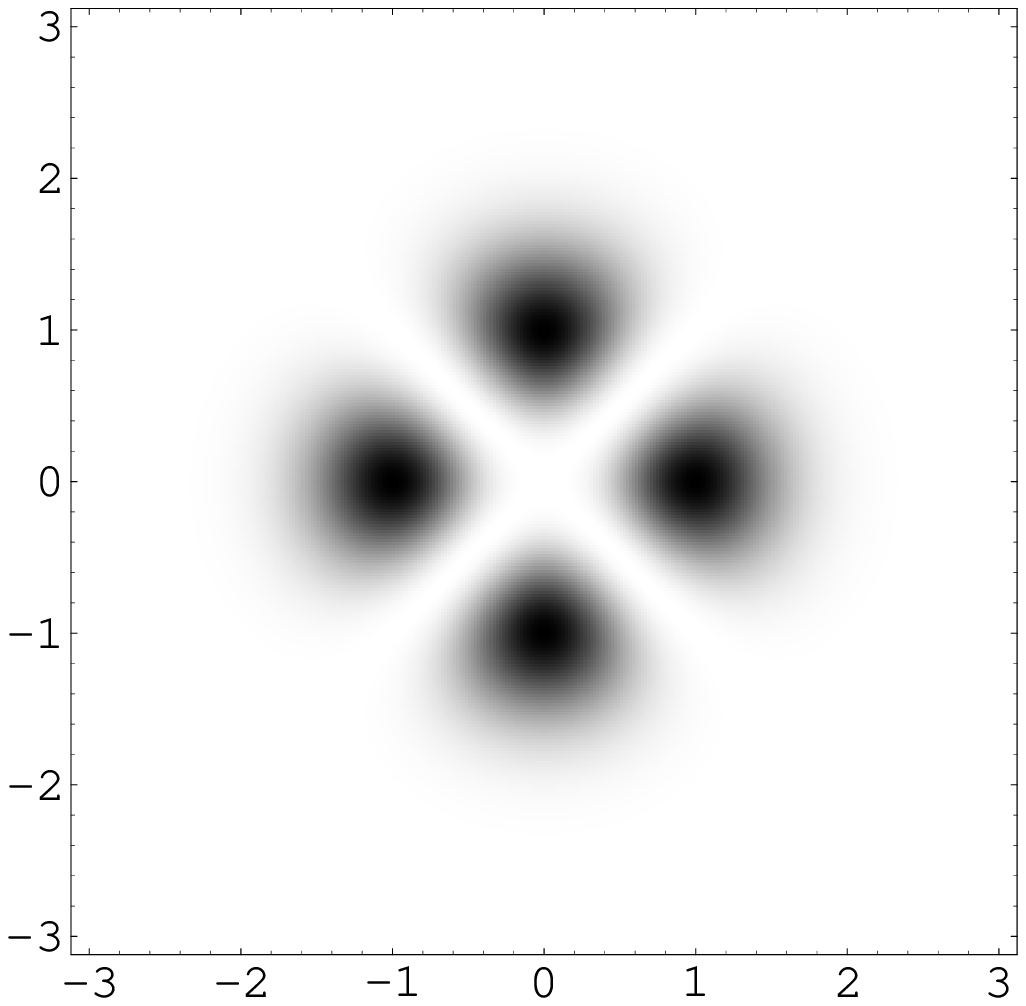}}
      \put(-2,21){$\scriptstyle v$}
     \put(21,-2){$\scriptstyle y$}
   \end{picture}     
      \caption{The marginal Wigner function $W_\mathrm{m}(y,v)$ for $N=2$.}\label{Wigner2}
        \end{figure}         
  \begin{figure}[htb]
    \setlength{\unitlength}{1mm}
     \begin{picture}(40,40) 
      \put(0,0){\includegraphics[scale=0.4]{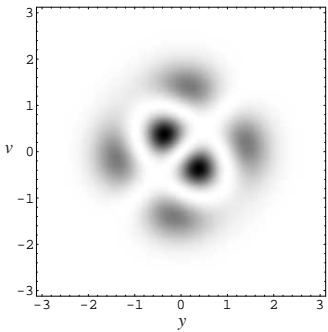}}
      \put(-2,21){$\scriptstyle v$}
     \put(21,-2){$\scriptstyle y$}
   \end{picture}          
      \caption{The marginal Wigner function $W_\mathrm{m}(y,v)$ for $N=3$.}\label{Wigner3}
        \end{figure} 
From the density plots in Fig.~\ref{Wigner1}, \ref{Wigner2}, and \ref{Wigner3}
we see that the probability densities become more symmetric the larger $N$
becomes, similar to the previous case for the marginals of the $Q$
function, but the interference structures are much more pronounced than
for the $Q$ function. Here we also obtain a vanishing correlation coefficient
$r$ for all $N>1$, from which we can infer that a nonlinear correlation measure
is necessary to describe these correlations. 
\par We conclude from the results of the first section that a set of parameters
         can always be found which violate the CH inequality in
Eq.~(\ref{CH}). Therefore N00N states show EPR correlations for any finite
$N$. The presented setup is not yet optimal but might be promising for
demonstrating EPR correlations of N00N states with low photon numbers $N$
experimentally. Although, the requirements for the overall detection efficiency for a loophole-free test of the CHSH Bell
inequality would be very large, i.e., $96\%$ for $N=1$ \cite{Bellini}.  
\par In the following section we are going to show that the test of other Bell-type
inequalities leads to a different result. 
\section{More Bell-type inequalities}\label{new_inequalities}
\par So far we have used the CH and the CHSH Bell inequalities defined in
Eq.~(\ref{CH}) and Eq.~(\ref{CHSH}). Other
Bell inequalities might be more suitable for a Bell test for a nonlocal experiment with N00N states. 
The CH Bell inequality is a specific inequality for four correlated events,
where at most two are intersected at the same time. Pitowsky \cite{Pitowsky}
derived all the Bell-type inequalities for three and four correlated events:
\begin{equation}
  0\le p_i-p_{ij}-p_{ik}+p_{jk}\,,\label{Bell-Wigner1}
\end{equation}
\begin{equation}
  p_i+p_j+p_k-p_{ij}-p_{ik}-p_{jk}\le 1\,,\label{Bell-Wigner2}
\end{equation}
\begin{equation}
  -1\le p_{ik}-p_{j\ell}+p_{i\ell}+p_{jk}-p_i-p_k\le 0\,,\label{Bell-CH}
\end{equation}
for any different $i,j,k,\ell$.
Eq.~(\ref{Bell-CH}) is the CH inequality. Eqs.~(\ref{Bell-Wigner1},\ref{Bell-Wigner2})
are inequalities in the so called Bell-Wigner polytope of three correlated
events, whereas Eq.~(\ref{Bell-CH}) belongs to the Clauser-Horne polytope
\cite{Pitowsky}. Later on Janssens et al. \cite{Fuzzy} explicitly constructed
inequalities for six correlated events where, as before, two are
intersected at the same time. We consider the following four:
 \begin{equation}
  p_i+p_j+p_k+p_\ell-p_{ij}-p_{ik}-p_{i\ell}-p_{jk}-p_{j\ell}-p_{k\ell}\le
  1,\label{six_2}
\end{equation}
\begin{equation}
  2p_i+2p_j+2p_k+2p_\ell-p_{ij}-p_{ik}-p_{i\ell}-p_{jk}-p_{j\ell}-p_{k\ell}\le 3,\label{six_3}
\end{equation}
 \begin{equation}0\le
  p_i-p_{ij}-p_{ik}-p_{i\ell}+p_{jk}+p_{j\ell}+p_{k\ell},\label{six_4}
\end{equation}
\begin{equation}
  p_i+p_j+p_k-2p_\ell-p_{ij}-p_{ik}+p_{i\ell}-p_{jk}+p_{j\ell}+p_{k\ell}\le 1.\\\label{six_5}
   \end{equation}
We investigate the amount of violation for
the inequalities in Eqs.~(\ref{six_2}-\ref{six_5}) for the simple on-off
detection scheme of section \ref{Intro} with the detection probabilities given
by Eqs.~(\ref{Qfunction1},\ref{Qfunction2},\ref{Qfunction3}). The probabilities in Eq.~(\ref{six_2}) are then replaced by
\begin{eqnarray}
  J_1=Q(\alpha)+Q(\beta)+Q(\gamma)+Q(\delta)-Q(\alpha,\beta)-Q(\alpha,\gamma)\nonumber\hspace{-2mm}\\{}-Q(\alpha,\delta)-Q(\beta,\gamma)-Q(\beta,\delta)-Q(\gamma,\delta)\label{six_21},
\end{eqnarray}
so that the inequality is given by $J_1\le 1$. We make the following
assignment $i\rightarrow \alpha$, $j\rightarrow \beta$, $k\rightarrow \gamma$,
and $\ell\rightarrow \delta$. The single-count probabilities
$Q(\alpha)$ can either be measured by Alice or by Bob. The joint probabilities are always
measured between Alice and Bob.  
\par
\begin{figure}[htb]
     \setlength{\unitlength}{1cm}  
         \begin{minipage}[t]{6cm}
         \begin{picture}(6,4)\includegraphics[scale=0.60]{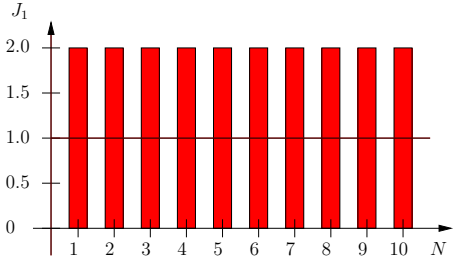}\end{picture}
           \caption{Violation of the inequality Eq.~(\ref{six_21}) as a function
             of $N$.}\label{display_21}
          \end{minipage}
          \end{figure}

 A maximization procedure carried out in the parameter space
$\alpha,\beta,\gamma,\delta$ leads to a constant violation of the inequality as shown in Fig.~\ref{display_21}.
This new result will be interpreted in more detail at the
end of this section together with the results from the remaining inequalities.
\par The probabilities in Eq.~(\ref{six_3}) can be rewritten in terms of the local oscillator amplitudes as well
\begin{eqnarray}
\lefteqn{J_2=2Q(\alpha)+2Q(\beta)+2Q(\gamma)+2Q(\delta)-Q(\alpha,\beta)}\nonumber\\
& & {}-Q(\alpha,\gamma)-Q(\alpha,\delta)-Q(\beta,\gamma)-Q(\beta,\delta)-Q(\gamma,\delta)\label{six_31},
\end{eqnarray}
where the inequality is then given by $J_2\le 3$.
A maximization procedure for the parameters in Eq.~(\ref{six_31}) shows a constant
violation of 4 for any $N$.
\par
\begin{figure}[htb]
     \setlength{\unitlength}{1cm}  
         \begin{minipage}[t]{6cm}
         \begin{picture}(6,4)\includegraphics[scale=0.60]{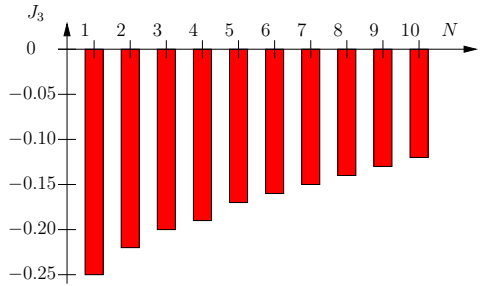}\end{picture}
           \caption{Violation of the inequality Eq.~(\ref{six_41}) as a function
             of $N$.}\label{display_41}
          \end{minipage}
          \end{figure}

\begin{figure}[htb]
     \setlength{\unitlength}{1cm}  
         \begin{minipage}[t]{6cm}
         \begin{picture}(6,4)\includegraphics[scale=0.60]{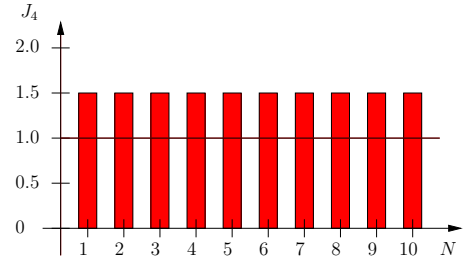}\end{picture}
           \caption{Violation of the inequality Eq.~(\ref{six_51}) as a function
             of $N$.}\label{display_51}
          \end{minipage}
          \end{figure}

 Finally the probabilities in Eq.~(\ref{six_4}) and Eq.~(\ref{six_5}) appear to be, in terms of
the complex parameters $\alpha,\beta,\gamma,\delta$,
\begin{eqnarray}
  J_3=Q(\alpha)-Q(\alpha,\beta)-Q(\alpha,\gamma)-Q(\alpha,\delta)\nonumber\\{}+Q(\beta,\gamma)+Q(\beta,\delta)+Q(\gamma,\delta),\label{six_41}
\end{eqnarray}
with the inequality $0\le J_3$. And
\begin{eqnarray}  
\lefteqn{J_4=Q(\alpha)+Q(\beta)+Q(\gamma)-2Q(\delta)-Q(\alpha,\beta)}\nonumber\\
& &{}-Q(\alpha,\gamma)+Q(\alpha,\delta)-Q(\beta,\gamma)+Q(\beta,\delta)+Q(\gamma,\delta)\label{six_51},
\end{eqnarray}
with the inequality $J_4\le 1$. 
Unlike the two previous cases we do not obtain a constant violation for
Eq.~(\ref{six_41}). Instead we attain a decreasing violation with the photon number $N$ as displayed in Fig.~\ref{display_41}.
So not all inequalities in the polytope of six correlated events can be violated
by a constant amount.
However the last inequality Eq.~(\ref{six_51}) is violated constantly again
with a value of $1.5$ as displayed in Fig.~\ref{display_51}. 
\par The Bell-type inequalities with six correlated events all show a stronger
violation than the CH and CHSH inequalities. We attain, except for one case, a
constant violation for any $N$. 
\par
\begin{figure}[htb]
\includegraphics[scale=0.38]{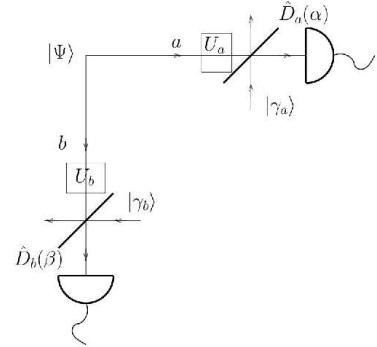}
\caption{Alice and Bob apply a local unitary operation on her/his mode $a$ and $b$.\label{homodyne_2}}
\end{figure}

 We expect that Bell inequalities exist which show a constant violation
because of the following argument. Let's
assume Alice and Bob can perform locally, a unitary transformation on her/his particle as
given by
\begin{eqnarray}
  U_{j}=|1\rangle_{j}\langle N|+|N\rangle_{j}\langle 1|+\sum_{n=0\atop n\not=1,N}^\infty |n\rangle_{j}\langle
   n|\,,
\end{eqnarray}
where $j=a,b$.
The combined application of their local unitary transformations transforms the one-photon
entangled state into an $N$-photon entangled N00N state
\begin{eqnarray}
  U_aU_b\frac{|1,0\rangle-|0,1\rangle}{\sqrt{2}}=\frac{|N,0\rangle-|0,N\rangle}{\sqrt{2}}\,,
\end{eqnarray}
(see also Fig.~\ref{homodyne_2}).
The fact that this local unitary operation exists tells us that there ought to
be a nonlocal measurement which acknowledges this fact.
Therefore the same amount of nonlocality should be obtained for the $N$-photon state as for
the one-photon entangled state. The fact that some of the Bell-tests do not show
this result means that these Bell-tests are not optimal. However the Bell-tests of the inequalities in
Eqs.~(\ref{six_21},\ref{six_31},\ref{six_51}) seem to be optimal for the N00N 
state since their outcome shows a constant violation for any $N$. We point out that
these Bell-type inequalities have two more joint probabilities than the CH and
CHSH Bell inequalities. The class of inequalities with six
joint probabilities seem to be more sensitive to the nonlocality in N00N states. From our results we also
infer, that for some applications, types of Bell inequalities other than the Clauser-Horne and
the Clauser-Horne-Shimony-Holt should be considered. It is, experimentally,
not more difficult to test these Bell inequalities; since one only needs to
measure the correlation functions for a few more parameter settings. The experimental setup does not need to be changed.  
\section{Conclusion}
\par We presented several Bell-tests for N00N states. In section \ref{on/off} a simple
on-off detection scheme together with the CH Bell inequality shows a violation for
any $N$ although the violation decreases as $N$ increases. In section \ref{parity} we
consider a correlated parity measurement together with the CHSH Bell
inequality. A violation is found only for $N=1$. In section \ref{new_inequalities} we
consider the simple on-off detection scheme but test Bell-type inequalities
with six joint probabilities. We then attain a violation that stays constant for any
$N$ and we show by a simple argument with local unitary operations that this is
to be expected for an optimal Bell-test with N00N states. 
If we use the violation of a Bell-type inequality as a measure of nonlocality
then N00N states contain the same amount of
nonlocality for any $N$. Despite this fact, using N00N states with large $N$ is advantageous for
applications like quantum imaging, metrology, and sensing, although the
improvement in the performance of these applications does not seem to be necessarily
related to the nonlocal properties of N00N states.
\par Finally we point out that it might be advantageous in many
experiments to also test the Bell-type inequalities in section
\ref{new_inequalities}, in addition to the CH or CHSH Bell inequalities. One gains more insight into the
nonlocal properties of the states under investigation, as shown by our
example.
\begin{acknowledgments}
C.F.W. and J.P.D. acknowledge the Hearne Institute for Theoretical Physics, the Disruptive
Technologies Office and the Army Research Office for
support. A.P.L. acknowledges the Australian Research Council and the Hearne
Institute for Theoretical Physics for support and
T. C. Ralph for valuable discussions. This work has also benefited from
helpful comments from H. V. Cable, W. Plick, M. M. Wilde, K. Jacobs,
D.~ H.~ Schiller, N. Sauer, and R. Kretschmer.\end{acknowledgments}

\end{document}